\documentclass[12pt]{article}
\usepackage{fleqn}
\usepackage{emlines2,bezier}
\usepackage{amssymb}
\author{S.S.Kokarev\thanks{e-mail: sergey@gw.yspu.yar.ru}\\
Yaroslavl State Pedagogical University,\\
150000, Yaroslavl, Respublikanskaya 108, r.409}
\title{Space-time as multidimensional elastic plate}
\date{}
\DeclareSymbolFont{AMSa}{U}{msa}{m}{n}
\DeclareMathSymbol\square{\mathord}{AMSa}{"03} \global\let\Box\square
\begin{document}

\maketitle \pagenumbering{arabic}
\begin{abstract}

It is suggested, that a curved 4-dimensional
space-time manifold is a strained
elastic plate in multidimensional embedding space-time. Its
thicknesses along ex\-tra\-di\-men\-si\-ons are much less than
4-di\-men\-si\-o\-nal
sizes.
Reduced 4-di\-men\-si\-o\-nal free energy density of the strained plate
in a weak strain case is similar to  GR Lagrangian density of a gravitational
field for the particular value of the Poisson coefficient
of the plate.
Dynamical equations of the theory are obtained by variation
of the multidimensional free energy over displacement vector components
$\xi^{A}$. In general case they are inhomogeneous bewave equations.

\end{abstract}


\section{Introduction}

In this  article we
suggest an alternative (to GR) approach to the
description of space-time dynamics. It is based on the
multidimensional formulation of elasticity theory (MET),
which is a direct
generalization of the standard 3-dimensional elasticity
theory.
Then the space-time continuum should be treated as
multidimensional physical medium with elastic properties.
Dimension of the observable space-time leads us to the concept
of {\it 4-dimensional plate.} It is a multidimensional body,
whose sizes along 4 dimensions are much greater then in other
ones. We suppouse, that these extradimensions  have certain
"thicknesses" (negligibly small) and consider
curvature of space-time as a manifestation of mechanical
straining of this multidimensional plate.
In some sence similar philosophy have been used
in \cite{clif}-\cite{pav}.

To compare our results with GR, we use embedding theory formalism
\cite{emb1}-\cite{emb5}.
In section  \ref{emb}
we present some usefull facts from the embedding theory.
The expression for the gravitational
field action in terms of displacement vector components derivatives
is obtained.

Section \ref{elast}
is devouted to the multidimensional generalization of the
standard elasticity theory
and to the theory of a thin multidimensional plate straining.
The formula  for the elastic free energy of the
multidimensional
plate, integrated over extra coordinates, is derived.
Then we compare variational functionals of
GR and MET.

In section \ref{dim} dimensional analysis is made
and some numerical evaluations
for Young's modulus of the multidimensional plate are obtained.

In section \ref{variating} we variate free energy functional
over  multidimensional  displacement vector components.
Mul\-ti\-di\-men\-si\-o\-nal ge\-ne\-ra\-li\-za\-ti\-on of Sophy-Zhermen
equation and boundary conditions are obtained.

In section \ref{shw}
Schwarzschild solution is analyzed from the viewpoint of MET.

\vspace{1cm}
\section{Embedding theory in deformation representation}\label{emb}

\begin{figure}[htb]
\centering
\unitlength=1.00mm
\special{em:linewidth 0.4pt}
\linethickness{0.4pt}
\begin{picture}(110.00,50.00)(0.00,40.00)
\bezier{312}(32.00,79.00)(67.00,95.00)(103.00,79.00)
\bezier{488}(32.00,79.00)(75.00,30.00)(103.00,79.00)
\put(92.00,64.00){\line(1,0){18.00}}
\put(110.00,64.00){\line(-2,-3){12.00}}
\put(98.00,46.00){\line(-1,0){77.00}}
\put(21.00,46.00){\line(3,4){13.67}}
\put(36.00,64.00){\line(1,0){2.00}}
\put(38.00,64.00){\line(-1,0){3.00}}
\put(35.00,64.00){\line(1,0){1.00}}
\put(35.00,64.00){\line(1,0){13.00}}
\put(53.00,53.00){\vector(0,1){13.00}}
\put(66.00,53.00){\vector(0,1){11.00}}
\put(79.00,54.00){\vector(0,1){10.00}}
\put(93.00,58.00){\vector(0,1){12.00}}
\put(93.00,70.00){\vector(0,0){0.00}}
\put(45.00,61.00){\vector(0,1){9.00}}
\put(50.00,77.00){\makebox(0,0)[cc]{$V_{1,3}$}}
\put(99.00,59.00){\makebox(0,0)[cc]{$M_{1,3}$}}
\put(33.00,51.00){\makebox(0,0)[cc]{$x^{\alpha}$}}
\put(58.00,61.00){\makebox(0,0)[cc]{$\xi^{A}$}}
\put(16.00,85.00){\makebox(0,0)[cc]{$M_{p+1,q+3}$}}
\bezier{308}(32.00,79.00)(69.00,64.00)(103.00,79.00)
\end{picture}
\caption{Embedding of $M_{1,3}$ and $V_{1,3}$ in $M_{p+1,q+3}$  in deformation representation}\label{embed}
\end{figure}
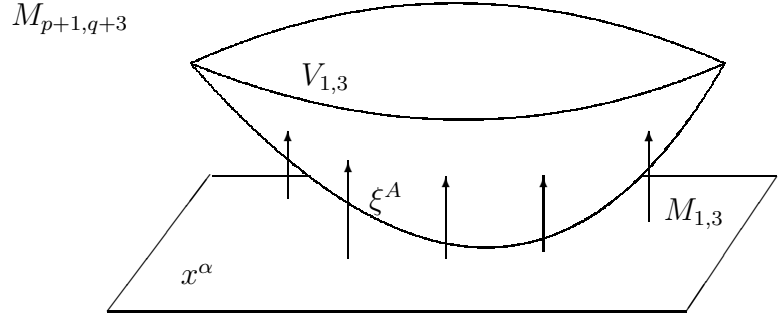

Let us consider
pseudoeuclidean
space ®f  $N+4$ dimensions, where  $p+1$ ones are time-like
and $q+3$ --- space-like: $Œ_{p+1,q+3}$ $(p+q=N)$.
Its metric can be defined as  $\eta_{AB}=\epsilon_{A}\delta_{AB},$
$A,B=\overline{1,N+4},\ \ \epsilon_{A}=\pm1$.
Hereafter big Latin letters
$A,B,C,\dots $ denote
tensor components with respect to the group
of rotation $O(p+1,q+3)$ in $M_{p+1,q+3}$,
Greek indices $\alpha,\nu,\lambda,\dots=0,1,2,3$ ---
are tensor components
with respect to a group of general coordinate transformations in
$M_{1,3}$ (and $V_{1,3}$)
and small Latin indices $m,n,\dots=\overline{1,N}$ ---
are tensor components in
subspace of $M_{p+1,q+3}$, orthogonal to $M_{1,3}$, where
4-dimensional plane $M_{1,3}$,
containes one time-like and three space-like directions
(fig.\ref{embed}).
Using the transformations from the isometry group of $M_{p+1,q+3}$
and renumerating Cartesian coordinates
in  $M_{p+1,q+3}$ we'll get the equation of this plane in the form:
\begin{equation}\label{pleq}
x^{m}=0, m=\overline{1,N}.
\end{equation}
Then four Cartesian coordinates $x^{\alpha}$ in  $M_{p+1,q+3}$
become Cartesian coordinates in
$M_{1,3}$, and induced metric in $M_{1,3}$
will be $\eta_{\alpha\nu}=diag\{+1,-1,-1,-1\}$.

Let $\xi^{A}(x^{\alpha})$ be smooth vector field defined on $M_{1,3}$,
orthogonal to $M_{1,3}$.
This vector field determines a deformation of $M_{1,3}$
into some curved Riemannian manifold $V_{1,3}$
(up to the parallel transition and rigid rotation in $M_{p+1,q+3}$).
The only nonzero components of $\xi^{A}$ are $\xi^{m}$
(due to (\ref{pleq}) and orthogonality to $M_{1,3}$).
Metric on $V_4$,
can be expressed in terms of  displacment vector derivatives over
coordinates $x^{\alpha}$:
\begin{equation}\label{metr}
ds^{2}=\eta_{AB}dx^{A}dx^{B}=\eta_{\alpha\nu}dx^{\alpha}dx^{\nu}+
\eta_{mn}\xi^{m}_{,\alpha}\xi^{n}_{,\nu}dx^{\alpha}dx^{\nu}=\eta_{AB}Y^{A}_{\alpha}Y^{B}_{\nu}dx^{\alpha}dx^{\nu}=
g_{\alpha\nu}dx^{\alpha}dx^{\nu},
\end{equation}
where
$Y^{A}_{\alpha}=\delta^{A}_{\alpha}+\xi^{A}_{,\alpha}$ ---
four tangent to  the $V_{1,3}$ vector fields.

Let's introduce
$N$ unit vectors $n^{A}_{m|}$ orthogonal to each other and to $V_{1,3}$:
\begin{equation}\label{ort}
\eta_{AB}n^{A}_{m|}n^{B}_{n|}=\epsilon_{m}\delta_{mn};
\end{equation}
\begin{equation}\label{ort1}
\eta_{AB}n^{A}_{m|}Y^{B}_{\alpha}=0,
\end{equation}
where vertical bar separates the number of the
vector from the coordinate index. Hereafter  we use notations of
\cite{eiz}.
Then at every point of $V_{1,3}$
tangent basis  $\vec Y$ can be supplemented to a basis in $M_{p+1,q+3}$.
The Cristoffel symbols in $V_{1,3}$ are the following:
\begin{equation}\label{Cris}
\Gamma^{\alpha}_{\nu\lambda}=Y^{A\alpha}Y_{A\nu,\lambda}.
\end{equation}
Raising and lowering of Greek indices is made with $g^{\alpha\nu}$ and
$g_{\alpha\nu}$, Latin indices --- with $\eta_{AB}$.

The covariant derivative
of the tangent vector
has the form:
\begin{equation}\label{cov}
 Y^{A}_{\alpha;\nu}=Y^{A}_{\alpha,\nu}-\Gamma_{\alpha\nu}^{\lambda}Y^{A}_{\lambda}=
Y^{A}_{\alpha,\nu}-Y^{B}_{\lambda }Y_{B\alpha,\nu}Y^{A\lambda}.
\end{equation}
This derivative is the tensor on $V_{1,3}$, and is the vector
in $M_{p+1,q+3}$,
orthogonal to $V_{1,3}$.
Due to this fact we can expand it with respect to $n^{A}_{m|}$:
\begin{equation}\label{2form}
Y^{A}_{\alpha;\nu}=\sum_{m}\epsilon_{m}\Omega_{m|\alpha\nu}n^{A}_{m|}.
\end{equation}
The symmetric  tensor $\Omega_{m|\alpha\nu}$ is multidimensional
generalization
of a second fun\-da\-men\-tal form
of a 2-dimensional surface in 3-dimensional space.
Scalar product of
(\ref{2form}) and $n_{s|A}$ gives (taking into account (\ref{ort}) and
(\ref{cov})):
\begin{equation}\label{2fd}
\Omega_{s|\alpha\nu}=n_{s|A}Y^{A}_{\alpha;\nu}=n_{s|A}Y^{A}_{\alpha,\nu}=
n_{s|m}\xi^{m}_{,\alpha,\nu}.
\end{equation}
Integrability condition for Eq. (\ref{2form}) has the form:
\begin{equation}\label{int}
2Y^{A}_{\alpha[;\nu;\lambda]}=R^{\sigma}_{\alpha\nu\lambda}Y^{A}_{\sigma}.
\end{equation}
We need only those conditions from
(\ref{int}), which give relation between curvature tensor
of inner geometry and the quadratic combinations
of $\Omega_{m|\alpha\nu}$:
\begin{equation}\label{gauss}
R_{\alpha\nu\lambda\sigma}=
\sum_{m=1}^{N}\epsilon_{m}
(\Omega_{m|\alpha\lambda}\Omega_{m|\nu\sigma}-
\Omega_{m|\alpha\sigma}\Omega_{m|\nu\lambda}),
\end{equation}
which are called {\it Gauss equations.}
From (\ref{gauss}) and (\ref{2fd}) one gets
\begin{equation}\label{riem}
R_{\alpha\nu\lambda\sigma}=
\sum_{m=1}^{N}
(\xi^{n}_{,\alpha,\lambda}\xi^{l}_{,\nu,\sigma}-
\xi^{n}_{,\alpha,\sigma}\xi^{l}_{,\nu,\lambda})
\epsilon_{m}n_{m|n}n_{m|l}=
(\xi^{n}_{,\alpha,\lambda}\xi^{l}_{,\nu,\sigma}-
\xi^{n}_{,\alpha,\sigma}\xi^{l}_{,\nu,\lambda})H_{nl},
\end{equation}
where $H_{nl}\stackrel{\rm def}{\equiv}\sum_{m=1}^{N}\epsilon_{m}n_{m|n}n_{m|l}$ ---
is the projector on orthogonal to
$V_{1,3}$ directions.
From (\ref{riem}) we can easily find
Ricci tensor and curvature scalar.

Thus
gravitational field action in GR can be
written in terms of embedding variables:
\begin{equation}\label{act}
S_{g}=-\frac{c^{3}}{16\pi G}\int R\sqrt{-g}d^{4}x=
-\frac{c^{3}}{16\pi G}\int\left\{
(\xi^{n\ \ \alpha}_{,\alpha,}\xi^{l\ \ \nu}_{,\nu,}-
\xi^{n}_{,\alpha,\nu}\xi^{l\alpha\ \ \nu}_{,\ \ ,})
H_{nl}\right\}\sqrt{-g}d^{4}x.
\end{equation}
It is cleare now that
embedding theory
allows us to rewrite
gravitational action in terms of derivatives of multidimensonal
displacement vector.

\section{Multidimentional elasticity theory\protect\\  and strain
of a multidimensional plate.}
\label{elast}

Let us consider some  multidimensional body in $M_{p+1,q+3}$ in unstrained state
and a Cartesian coordinates system rigidly tied with it.
Every point of the strained body displaces on
some vector $\vec \xi$, which is a function of coordinates.
The square of the length between  two neighbour points
of the strained body is given by the following expression
\begin{equation}
dl'^{2}=dl^{2}+2\xi_{AB}dx^{A}dx^{B}
\end{equation}
where symmetric tensor
\begin{equation}\label{def}
\xi_{AB}=\frac{1}{2}\left(
\frac{\partial \xi_{A}}{\partial x^{B}}+
\frac{\partial \xi_{B}}{\partial x^{A}}+
\frac{\partial \xi_{M}}{\partial x^{A}}
\frac{\partial \xi^{M}}{\partial x^{B}}\right),
\end{equation}
is called the {\it strain tensor} (see for example \cite{land}).
Quadratic terms in (\ref{def})
are negligibly small for small strains and can be ommited.

Inner forces, induced by the deformation, act on
a surface of the separated volume only and, consequently,
the resulting force, acting on the chosen volume, has the
form:
\begin{equation}\label{forth}
F_{A}=\int\limits_{V}f_{A}dV=\int\limits_{V}\frac{\partial\sigma_{AB}}{\partial x^{B}}dV=
\oint\limits_{\partial V}\sigma_{AB}ds^{B} ,
\end{equation}
where $f_{A}$ --- $A$ -th component of the force volume density, the $\sigma_{AB}$ ---
{\it stress tensor} and $f_{A}=\sigma_{AB,}^{\ \ \ B}$,
$\partial V$ --- is the
hypersurface, bounding separated volume.
Tensor $\sigma_{AB}$
is not uniquelly determined from (\ref{forth})
and it can always be written in the symmetric form.

Equilibrium equations in the absence of an external
forces are:
\begin{equation}
\frac{\partial\sigma_{AB}}{\partial x^{B}}=0.
\end{equation}
The boundary conditions of equilibrium equations in elasticity theory
has the form:
\begin{equation}\label{bound}
\sigma_{AB}n^{B}=P_{A},
\end{equation}
where $P_{A}$ --- hypersurface force density,
$\vec n$ --- unit vector normal to the surface.

From the multidimensional generalization of the first law of
thermodynamics
\[
dF=-SdT+\sigma_{AB}d\xi^{AB}
\]
one can find:
\begin{equation}\label{tens}
\sigma_{AB}=\left(\frac{\partial F}{\partial \xi^{AB}}\right)_{T}
\end{equation}
where $\delta A=-\sigma_{AB}\delta\xi^{AB}$ is the
work of inner elastic forces, $F=E-TS$ ---
multidimensional free energy density, $E$ ---
multidimensional internal energy density,
$T$ --- multidimensional absolute temperature, $S$ ---
multidimensional enthropy density.

Equations of static elasticity theory can be obtained by
variation of $F$ over
displacement vector components.
For the isothermic weak strain case free energy can be expressed
in terms of strain tensor components:
\begin{equation}\label{free}
F=\frac{\lambda}{2}(\xi_{A}^{A})^{2}+\mu \xi_{AB}\xi^{AB}=
\mu\left(\xi_{AB}-\frac{1}{n}\eta_{AB}\xi_{C}^{C}\right)^{2}+
\frac{K}{2}(\xi_{C}^{C})^{2} =
\end{equation}
\[
\frac{E}{2(1+\sigma)}\left(\xi_{AB}\xi^{AB}+
\frac{\sigma}{1-(n-1)\sigma}(\xi_{C}^{C})^{2}\right),
\]
where $\lambda$ and $\mu$ are called {\it Lame coefficients},\\
$K=\lambda+(2/n)\mu$ --- {\it stretch modulus}, $\mu$ --- {\it shear
modulus,}\\
\[
E=\frac{n^{2}K\mu}{(1/2)n(n-1)K+\mu},\ \
\sigma=\frac{1}{2}\frac{(nK-2\mu)}{[(1/2)n(n-1)K+\mu]}
\]
--- {\it Young's modulus and Poisson coefficient} respectively.
Here $n$ --- is the total number of dimensions --- $N+4$.
To clarify physical meaning
of Young's modulus and Poisson coefficient let us consider
a simple stretch of a bar along his axe of symmetry
(axe $x^{1}$). Then $\xi_{11}=p/E,\ \xi_{AB}=-\epsilon_{1}\eta_{AB}\sigma\xi_{11},
\ A,B\neq 1$ ,
where $p$ --- stretch force, acting on the unit hyphersurface
of the bar end.

Taking (\ref{tens}) and (\ref{free})
one can obtain expression for a stress tensor in linear elasticity
theory:
\begin{equation}\label{ltens}
\sigma_{AB}=2\mu\xi_{AB}+\lambda\eta_{AB}\xi_{C}^{C}=
\frac{E}{1+\sigma}\left(\xi_{AB}
+\frac{\sigma}{1-(n-1)\sigma}\eta_{AB}\xi_{C}^{C}\right).
\end{equation}

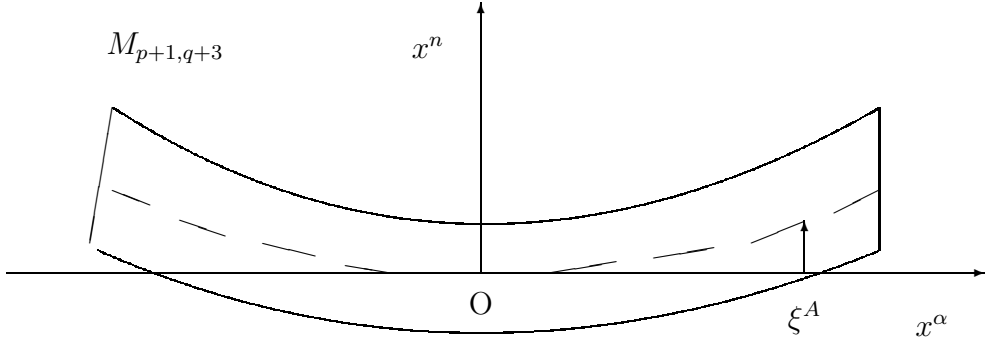
\begin{figure}[htb]
\centering
\unitlength=1mm
\special{em:linewidth 0.4pt}
\linethickness{0.4pt}
\begin{picture}(139.00,60.00)(0.00,50.00)
\bezier{476}(23.00,92.00)(71.00,61.00)(125.00,92.00)
\bezier{452}(21.00,73.00)(71.00,51.00)(125.00,73.00)
\put(23.00,92.00){\line(-1,-6){3.00}}
\put(125.00,92.00){\line(0,-1){19.00}}
\put(9.00,70.00){\vector(1,0){130.00}}
\put(72.00,70.00){\vector(0,1){36.00}}
\put(23.00,81.00){\line(5,-2){5.00}}
\put(32.00,77.00){\line(3,-1){6.00}}
\put(43.00,74.00){\line(4,-1){7.00}}
\put(54.00,71.00){\line(6,-1){6.00}}
\put(81.00,70.00){\line(6,1){9.00}}
\put(95.00,72.00){\line(6,1){8.00}}
\put(108.00,74.00){\line(5,2){7.00}}
\put(119.00,78.00){\line(2,1){6.00}}
\put(132.00,63.00){\makebox(0,0)[cc]{$x^{\alpha}$}}
\put(65.00,100.00){\makebox(0,0)[cc]{$x^{n}$}}
\put(72.00,66.00){\makebox(0,0)[cc]{O}}
\put(115.00,70.00){\vector(0,1){7.00}}
\put(115.00,64.00){\makebox(0,0)[cc]{$\xi^{A}$}}
\put(30.00,100.00){\makebox(0,0)[cc]{$M_{p+1,q+3}$}}
\end{picture}
\caption{Straining of a thin plate}\label{defor}
\end{figure}

Let's consider a weak strain of a thin 4-dimensional plate.
Let $h_m$  be thickness of the plate in the $m$-th
extradimension orthogonal to $M_{1,3}$. Suppouse, that the
middle (in thickness) plane of the unstrained plate coinsides
with $M_{1,3}$-plane and $V_{1,3}$ is the surface of weakly strained
plate. The middle surface in the case of weak straines is usually
called {\it newtral surface} because tangent stresses on it are equal to
zero. Putting the origin of (p+1,q+3)-dimensional Cartesian
coordinate system on the newtral surface (Fig.\ref{defor})
one can treate $\xi^A$ vector field as the displacement vector
of newtral surface points in the orthogonal to $M_{1,3}$ directions.
As in standard 3-dimensional elasticity theory we can put
$P_{A}$ in (\ref{bound}) to zero,
(because $P_{A}$ is negligibly small in comparison with
$\sigma_{AB}$) and consider $n_{m|}^{A}$ as orthogonal vectors
to $M_{1,3}$-plane (because of smallness of plate bending).
Thus instead of (\ref{bound}) we have
\begin{equation}\label{msurf}
\sigma_{nA}=0.
\end{equation}
From (\ref{ltens})
we get the following system of equations for
the components of weak strain tensor
$\xi_{AB}=(1/2)(\xi_{A,B}+\xi_{B,A})$
\[
\xi_{mn}=\xi_{\alpha n}=0,\ (m\neq n);\  2\mu\xi_{mm}+\lambda\xi_{A}^{A}\epsilon_{m}=0.
\]
Its solution is
\[
\xi_{\alpha}=-x^{m}\xi_{m,\alpha}
\]
Then for $\xi_{AB}$ we obtain $(\mu\neq 0,\ \ N\lambda+2\mu\neq 0)$:
\begin{equation}\label{sol}
\xi_{\alpha n}=\xi_{mn}=0;\ \  \xi_{\alpha\beta}=-\xi_{m,\alpha,\beta}x^{m};
\end{equation}
\[
\xi_{mm}=-\epsilon_{m}\frac{\lambda S}{N\lambda+2\mu};\ \
S=\xi_{\alpha}^{\alpha}=-x^{m}\xi_{m,\alpha,}^{\ \ \ \alpha}.
\]
Substituting (\ref{sol}) into (\ref{free}) we get
\begin{equation}\label{freepl}
F=\mu x^{m}x^{n}\left\{\xi_{m,\alpha,\nu}\xi_{n,\ ,}^{\ \alpha\ \nu}+
f\xi_{m,\alpha,}^{\ \ \alpha}\xi_{n,\nu,}^{\ \ \nu}\right\},
\end{equation}
where $f=\lambda/(N\lambda+2\mu)$.

To obtain expression for total free energy of the plate one should integrate
(\ref{freepl}) over its $N+4$-dimensional volume:
\begin{equation}\label{freegen}
F_{\rm pl}=\frac{\mu H_{N}h_{m}^{2}}{12}\eta^{mn}\int \sqrt{-g}d^{4}x
\left\{\xi_{m,\alpha,\nu}\xi_{n,\ ,}^{\ \alpha\ \nu}+
f\xi_{m,\alpha,}^{\ \ \alpha}\xi_{n,\nu,}^{\ \
\nu}\right\}.
\end{equation}
where
\[
\delta^{mn}H_{N}\frac{h_{m}^{2}}{12}=
\delta^{mn}(\prod_{n=1}^{N}h_{n})\frac{h_{m}^{2}}{12}=
\prod\limits_{p=1}^{N}\int\limits_{-h_{p}/2}^{h_{p}/2}dx^{p}x^{m}x^{n},
\]
$H_{N}$ is the  product of all thicknesses of the plate,
$\delta^{mn}$ --- Kronecker symbol, $\eta^{mn}$ ---
are orthogonal to $M_{1,3}$ components of $\eta_{AB}$.

Comparing
(\ref{freegen})
and (\ref{act})
we may note their remarkable similarity.
Let's find the conditions under which (\ref{freegen})
and (\ref{act}) will be identical:
\begin{enumerate}
\item
The coefficient  $f$ in (\ref{freegen}) must be equal to $-1$:
\begin{equation}\label{c1}
f=-1.
\end{equation}
Taking into account expression for generalized
mul\-ti\-di\-men\-si\-o\-nal Lame coefficients
\begin{equation}\label{lame}
K=\frac{E}{(1-(n-1)\sigma)n};\ \ \mu=\frac{E}{2(1+\sigma)};\ \
\lambda=\frac{E\sigma}{(1+\sigma)(1-(n-1)\sigma)};
\end{equation}
we get the first condition
\begin{equation}\label{poiss}
\sigma=\frac{1}{n-N-2}
\end{equation}
and for $n=N+4$ one can easily obtain $\sigma=1/2$.
\item
In the case of weak deformation we can put in (\ref{act}):
$$
H_{mn}\approx\eta_{mn}
$$
It yields the second conditions
\begin{equation}\label{c2}
h_{m}=h,
\end{equation}
i.e. plate must have equal thicknesses in all extradimensions.
Easy to see, that last condition is connected with norming
conditions (\ref{ort}), and simple renormalization of vector
$n_{m|}^{A}$:
\[
\eta_{AB}n_{m|}^{A}n_{n|}^{B}=h_{m}^{2}\varepsilon_{m}\delta_{mn}
\]
gives us an opportunity  to formulate the theory without
the condition (\ref{c2})
(we use (\ref{ort}) to save the standard form of equations of
embedding theory).

This comparison leads us directly to the idea, that
{\it our observable macroscopic 4-di\-men\-si\-o\-nal space-time
can be described as the thin multidimensional plate,
"made" of some specific material. Its thicknesses are much
less, then its 4-dimensional sizes and its free energy
functional is defined by (\ref{freegen}).}
\end{enumerate}

\section{Dimensional analysis.}\label{dim}

Equating (\ref{act}) and (\ref{freegen}) and introducing constant
$A$  we get (with the conditions (\ref{c1}) and (\ref{c2})):
\begin{equation}\label{eq}
\frac{c^{3}}{16\pi G}\equiv A\frac{\mu h^{N+2}}{12}
\end{equation}
where $A$ is measured in seconds.
There are only two combinations of fundamental constants
$G,\hbar,c,e,h$,
($h$ --- thickness of the plate), which have appropriate
dimension "second":
$$
\tau=h/c\  \mbox{and}\  \tau_{\rm
pl}=\sqrt{G\hbar/c^{5}}\approx10^{-44}s,
$$
where $\tau_{pl}$ is Planck time.
From $\pi$ -theorem
in dimension theory (see for example \cite{sed})
\begin{equation}\label{pi}
A=\varphi_{1}(\frac{e^{2}}{\hbar c},\frac{\tau}{\tau_{pl}})\tau+
\varphi_{2}(\frac{e^{2}}{\hbar c}, \frac{\tau}{\tau_{pl}})\tau_{pl},
\end{equation}
where $\varphi_{1},\varphi_{2}$ --- two arbitrary functions
of two dimensionless combination of fun\-da\-men\-tal constants.
In the simplest case of  multiplicative dependence
of
$A$ on the fundamental constants we have
\begin{equation}\label{eq1}
Eh^{N+S+3}\sim c^{4-(3/2)S-P}G^{S/2-1}\hbar^{S/2-P}e^{2P},
\end{equation}
Here $S,P$ --- arbitrary real numbers.
From (\ref{eq1}) one can get a wide class of relations
between elastic constants and  fundamental ones.
Let us consider two most interesting cases:

a) Assuming, that
extradimensions have sizes of the Planck length
$h_{pl}=\sqrt{G\hbar/c^{3}}\approx10^{-33}$cm
(as in Kaluza-Klein theories), we obtain
\begin{equation}\label{pl}
E\sim c^{3/2N-P+17/2}G^{-5/2-N/2}\hbar^{-P-N/2-3/2}e^{2P}\sim10^{144+35N-3P}\mbox{Pa}
\end{equation}
We see,  that
space-time plate in this case possess by a huge stiffness
(compare with chromium $\sim10^{12}$Pa).
It gives natural explanation of observable "flatness"
of space-time in local regions.

b) For different values of parametres  $P$ and  $S$
one gets:

1) If $P=0,\ S=0$   then constants $e$ and $\hbar$ vanish
\[
Eh^{N+3}\sim\frac{c^{4}}{G}.
\]
This case is the most realistic
(due to the absence of $e$ and $\hbar$ in the classical approach to
gravity) and connected with old Sacharov's idea \cite{sach}
about possible elasticity of space-time.

2) If $S=2,\ P=1$ then constants $G,\hbar,c$ vanish
\[
Eh^{N+5}\sim e^{2};
\]

3) If $S=-N-3,\ P=-(N+3)/2$\  then $h$ and  $\hbar$ vanish
\[
E\sim\frac{ec^{2}}{\sqrt{G}}\left/\left(\frac{\sqrt{G}e}
{c^{2}}\right)^{N+4}\sim
10^{151+36N}\mbox{Pa}\right.;
\]

4) If $P=0,\ S=2$\   then $e$ and $G$ vanish
\[
Eh^{N+5}\sim c\hbar;
\]

5) If $P=0,\ S=8/3$\  then $e$ and  $á$ vanish
\[
Eh^{N+17/3}\sim G^{1/3}\hbar^{4/3};
\]

6) If $P=0,\ S=-N-3$\  then $e$ and $h$ vanish
\[
E\sim\sqrt{\frac{\hbar c^{5}}{G}}
\left/\left(\frac{G\hbar}{c^{3}}\right)^{(N+4)/2}\sim
10^{144+34N}\mbox{Pa};\right.
\]

7) Constants $G$ and  $h$ cannot be excluded from (\ref{eq1}) simultaneously;

8) If $S=-N-3,\ P=4+(3/2)(N+3)$  then $c$ and $h$ vanish
\[
E\sim\left(\frac{e^{5}}{\sqrt{G}\hbar^{2}}\right)/\left(\frac{\sqrt{G}\hbar^{2}}{e^{3}}\right)^{N+4}\sim
10^{118+30N}\mbox{Pa}.
\]

Presented analysis allows us to
formulate the second  hyphothesis, which developes the first one:
{\it
Space-time plate is a very stiff body in mul\-ti\-di\-men\-si\-o\-nal
space-time.
The Newtonian gravitational constant and
elastic constants of the plate are related by algebraic equation
(in the most realistic case: $Eh^{N+3}\sim 1/\ae$.)}

\section{Equilibrium equation and boundary
conditions}\label{variating}

Assume (as in standard 3-dimensional approach), that the volume of
the plate does not change when bending.
Then $\sqrt{-g}\approx1$.
Varying functional $F$ over $\xi^m$ we have:
\[
\delta F=\frac{\mu H_{N}h_{m}^{2}}{12}\eta_{mn}\delta
\int d^{4}x\left\{\xi_{,\alpha,\nu}^{m}\xi^{n \alpha\ \nu}_{,\ ,}+f\Box\xi^{m}
\Box\xi^{n}\right\}=
\]
\[
\frac{\mu H_{N}h_{m}^{2}}{12}\eta_{mn}2\delta\int
d^{4}x\left\{\frac{f+1}{2}\Box\xi^{m}\Box\xi^{n}+\sum_{\alpha\neq\nu,\ \alpha<\nu}
(\xi^{m}_{,\alpha,\nu}\xi^{n \alpha\ \nu}_{,\ ,}-\xi^{m\ \ \alpha}_{,\alpha,}\xi^{n\ \ \nu}_{,\nu,})\right\}=
\]
\begin{equation}\label{var}
D_{m}\eta_{mn}\delta\int d^{4}x\left\{\frac{1}{2}\Box\xi^{m}\Box\xi^{n}+
\frac{1}{f+1}\sum_{\alpha\neq\nu,\ \alpha<\nu}
(\xi^{m}_{,\alpha,\nu}\xi^{n \alpha\ \nu}_{,\ ,}-
\xi^{m\ \ \alpha}_{,\alpha,}\xi^{n\ \ \nu}_{,\nu,})\right\},
\end{equation}
where  $\Box\equiv\partial^{\alpha}\partial_{\alpha}$ --- wave operator,
\[
D_{m}=\frac{\mu H_{N}h_{m}^{2}(f+1)}{6}=\frac{EH_{N}h_{m}^{2}}{12(1+\sigma)}
\cdot\frac{1+\sigma(N-n+2)}{1+\sigma(N-n+1)}
\]
is the cylindrical stiffness factor of the plate in $m$-th
extradimension\footnote{For $n=3$ $N=1$ we have standard formulae
(\cite{land}).}.
Variation of the first term in (\ref{var}) gives
\begin{equation}
\delta\frac{1}{2}\int d^{4}x(\Box\xi^{m})^{2}=\int_{\Sigma}d^{4}x\Box^{2}\xi^{m}\delta\xi^{m}+
\oint_{\partial\Sigma}d^{3}S^{\alpha}\Box\xi^{m}\delta\xi^{m}_{,\alpha}-\oint_{\partial\Sigma}
d^{3}S^{\alpha}\Box\xi^{m}_{,\alpha}\delta\xi^{m},
\end{equation}
where $\Sigma$ is 4-dimensional plate surface, $\partial\Sigma$ is
3-dimensional bound of $\Sigma$.
Variation of the second term in (\ref{var}) yields:
\begin{equation}
\delta\int d^{4}x\sum_{\alpha\neq\nu,\ \alpha<\nu}
(\xi^{m}_{,\alpha,\nu}\xi^{m \alpha\ \nu}_{,\ ,}-
\xi^{m\ \ \alpha}_{,\alpha ,}\xi^{m\ \ \nu}_{,\nu,})=
\oint_{\partial\Sigma}d^3S^{\alpha}w_{\alpha}^{m},
\end{equation}
where $w_{\alpha}^{m}=\delta\xi^{m}_{,\lambda}\xi^{m\ \lambda}_{,\alpha,}-
\delta\xi^{m}_{,\alpha}\xi^{m\ \lambda}_{,\lambda,}$.
Variation of potential energy of the plate (due to applied bending
forces)
is $\delta U=-\int d^{4}xP_{n}\delta\xi^{n}$.
Finally we obtain:
\begin{equation}\label{varf}
\delta F_{\rm pl}+\delta U=\int_{\Sigma}d^{4}x(D_{m}\Box^{2}\xi_{m}-P_{m})\delta\xi^{m}-
D_{m}\oint_{\partial\Sigma}d^{3}S^{\alpha}\Box\xi_{m,\alpha}\delta\xi^{m}+
\end{equation}
\[
\frac{D_{m}}{1+f}\oint_{\partial\Sigma}d^{3}S^{\alpha}
(f\eta_{\alpha\lambda}\Box\xi_{m}+\xi_{m,\alpha,\lambda})\delta\xi^{m\lambda}_{,}
\]
Equating (\ref{varf}) to zero we get equilibrium equations
\begin{equation}\label{sg}
D_{m}\Box^{2}\xi^{m}=P^{m},
\end{equation}
where there is no summation over $m$.
Eqs. (\ref{sg}) are multidimensional generalization
of 3-D Sophy-Zhermen equilibrium equation.
Integrals over $\partial\Sigma$
give the boundary conditions for displacement vector and its
derivatives.
If all
$D_{m}=0$, then
we get GR theory.
In terms of $\xi^m$ functional $S_{g}$ (in linear approximation)
is a full divergence. Thus only boundary integrals do not vanish in
(\ref{varf}). It means, that there is no restrictions on (small)
displacement within $\partial\Sigma$:
any small bending of the plate is admissible if it meets
boundary conditions. The similar situation for ordinary 2-D plate
in standard 3-D elasticity theory can be obtained in unphysical limit:
$\sigma\to\infty.$

\section{Strained plate for Schwarzschild metric}\label{shw}

First let us prove equivalence of deformation representation
of embedding theory and standard representation of \cite{eiz}.
Let the set of functions
$\varphi^{A}(\zeta^{\alpha})$,
defining isometric embedding of $g_{\alpha\nu}$ in
$M_{p+1,q+3}$, is given, i.e:
\begin{equation}\label{metric}
ds^{2}=\eta_{AB}dx^{A}dx^{B}=\eta_{AB}\varphi^{A}_{,\alpha}\varphi^{B}_{,\nu}d\zeta^{\alpha}d\zeta^{\nu}=
g_{\alpha\nu}d\zeta^{\alpha}d\zeta^{\nu}.
\end{equation}
where $\zeta^{\alpha}$ --- arbitrary 4-dimensional coordinates.
Let four functions $\varphi^{\alpha}$ (from the set $\varphi^A$)
be equal to new 4-D coordinates:
\begin{equation}\label{tr}
\varphi^{\alpha}\equiv x^{\alpha}.
\end{equation}
Here (without loosing of generality) $\varphi^0$
corresponds to time-like dimension of $M_{p+1,q+3}$
and $\varphi^1,\varphi^2,\varphi^3$ - to space-like ones.
Besides we assume, that Jacobian
$D\varphi/D\zeta\neq0$.
Thus one can solve eq. (\ref{tr}) with respect to $\zeta$:
\begin{equation}\label{itr}
\zeta^{\alpha}=\zeta^{\alpha}(x).
\end{equation}
Substituting  (\ref{itr}) into the rest functions $\varphi^{m}$
we get:
\begin{equation}\label{f}
\varphi^m=\varphi^m(\zeta(x))=\bar{\varphi}^{m}(x).
\end{equation}
Then metric (\ref{metric})   takes the form:
\begin{equation}\label{metrd}
ds^{2}=\eta_{\alpha\nu}dx^{\alpha}dx^{\nu}+
\eta_{mn}\bar{\varphi}^{m}_{,\alpha}\bar{\varphi}^{n}_{,\nu}dx^{\alpha}dx^{\nu}.
\end{equation}
Comparing (\ref{metrd}) and (\ref{metr})
we find that (\ref{metrd}) is written in deformation representation
with functions
$\bar{\varphi}^{m}(x)$ as components of displacement vector.

It is clear, that this representation is not unique,
because of: 1) uncertainity of an isometric embedding and 2)
freedom in choosing of functions $\varphi^{\alpha}$ as 4-dimensional coordinates.

It is well known \cite{ros}, that  Schwarzschild metric:
\[
ds^{2}=\left(1-\frac{r_{g}}{r}\right)dt^{2}-\frac{dr^{2}}{\displaystyle 1-\frac{r_{g}}{r}}-
r^{2}d\Omega^{2}
\]
admits isometric
embedding of class 2 in $M_{2,4}$ with
the following embedding functions:
\begin{equation}
X^{1}=\sqrt{1-\frac{r_{g}}{r}}\cos t;\ X^{2}=\sqrt{1-\frac{r_{g}}{r}}\sin t;\
X^{3}=f(r);
\end{equation}
\[
X^{4}=r\sin\theta\cos\varphi;\ X^{5}=r\sin\theta\sin\varphi;\ X^{6}=r\cos\theta,
\]
where  $f(r)$ may be found from the equation: $f'^{2}=
\left(r_{g}^{2}/4r^{4}+r_{g}/r\right)/\left(1-r_{g}/r\right)$.
Metric in $M_{2,4}$ has the form: $\eta_{AB}=diag(+,+,-,-,-,-)$.

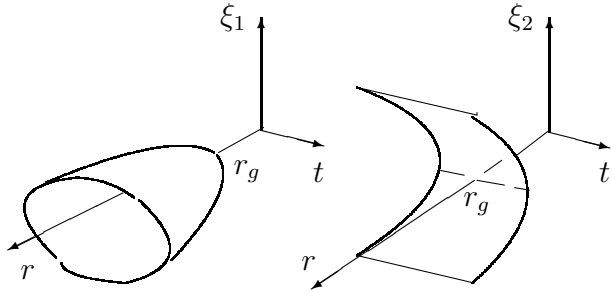
\begin{figure}[htb]
\unitlength=1mm
\special{em:linewidth 0.4pt}
\linethickness{0.4pt}
\begin{picture}(83.33,38.67)
\bezier{500}(12.33,17.33)(0.00,16.00)(9.67,6.67)
\bezier{500}(20.67,14.33)(29.67,6.00)(19.00,3.33)
\bezier{500}(10.33,6.00)(10.67,4.00)(18.67,3.33)
\bezier{500}(12.67,17.33)(17.00,17.00)(20.00,14.67)
\bezier{500}(8.33,16.67)(25.00,23.67)(30.67,20.67)
\bezier{500}(31.00,20.33)(34.33,16.33)(25.00,6.33)
\emline{31.33}{20.67}{1}{37.33}{24.00}{2}
\put(37.00,23.67){\vector(0,1){15.00}}
\put(37.00,23.67){\vector(4,-1){8.33}}
\put(18.67,15.33){\vector(-2,-1){15.33}}
\put(35.00,38.67){\makebox(0,0)[rc]{$\xi_1$}}
\put(44.67,19.33){\makebox(0,0)[ct]{$t$}}
\put(5.00,5.67){\makebox(0,0)[lt]{$r$}}
\put(33.33,19.67){\makebox(0,0)[lt]{$r_g$}}
\bezier{500}(49.67,7.00)(71.67,21.00)(49.67,29.33)
\bezier{500}(65.00,3.33)(79.67,16.33)(65.00,25.33)
\emline{49.67}{7.00}{3}{65.33}{3.33}{4}
\emline{49.67}{29.33}{5}{65.67}{25.67}{6}
\emline{65.67}{25.67}{7}{65.67}{26.00}{8}
\emline{60.67}{18.33}{9}{64.00}{17.67}{10}
\emline{65.67}{17.00}{11}{69.33}{16.33}{12}
\emline{70.67}{16.00}{13}{72.67}{15.67}{14}
\put(65.33,17.00){\vector(-3,-2){21.67}}
\emline{71.00}{20.33}{15}{75.33}{23.67}{16}
\put(75.33,23.33){\vector(0,1){15.33}}
\put(75.33,23.33){\vector(4,-1){8.00}}
\put(43.33,5.00){\makebox(0,0)[cb]{$r$}}
\emline{66.67}{18.00}{17}{69.00}{19.67}{18}
\put(65.67,15.33){\makebox(0,0)[ct]{$r_g$}}
\put(82.67,19.33){\makebox(0,0)[ct]{$t$}}
\put(73.33,38.67){\makebox(0,0)[rc]{$\xi_2$}}
\end{picture}
\caption{Strained 4-dimensional plate for Schwarzschild solution}\label{shwarc}
\end{figure}

Let  $(X^{1},X^{4},X^{5},X^{6})$ be coordinates on $M_{1,3}$,
then $X^{2}\equiv\xi^{1},\ X^{3}\equiv\xi^{2}$
take the form:
\begin{equation}
\xi^{1}=\sqrt{1-\frac{r_{g}}{r}-t^{2}};\ \xi^{2}=f(r).
\end{equation}
The corresponding strained plate is shown on figures \ref{shwarc}.
The first figure shows the form of the strained plate
in $\xi^{1}$-dimension: it is a paraboloid of revolution with
respect to $r$-axe. Second figure shows the form of the plate
in $\xi^{2}$-dimension: it is an infinite cylindrical surface.
Every point of this surfaces is a unit sphere in 3-dimensional
space section.
There is no 4-dimensional space-time for
$r<r_{g}$, so such 4-dimensional coordinate system is incomplit.

Since the Schwarzschild metric is the solution of Einstein equation
then variational principle (\ref{varf}) gives boundary conditions
only:
\begin{equation}\label{shbd}
\frac{EH_{N}h_{m}^{2}}{12(1+\sigma)}
\oint_{\partial\Sigma}d^{3}S^{\alpha}
(\Box\xi_{m}\delta^{\lambda}_{\alpha}-
\xi_{m,\alpha,}^{\ \ \ \lambda})
\delta\xi^{m}_{,\lambda}=0,
\end{equation}
For $n=6$ the coefficient in (\ref{shbd})
is nonzero, then  the integral  equals zero.
If plate has free or simply supported ends,
then variations of displacement vector and its derivatives
are independent and
\begin{equation}\label{shbd1}
\Box\xi_{m}\delta^{\lambda}_{\alpha}-
\xi_{m,\alpha,}^{\ \ \ \lambda}=0
\end{equation}
Direct calculation shows, that (\ref{shbd1}) is false and we can conclude,
that if $n=6$, then the bending plate (corresponding to Schwarzschild
solution) must have pinned ends at infinity (as in GR).

It is clear from Fig.\ref{shwarc}, that
the plate, corresponding to Scwarcshild metric, is bent
considerably,
so the linear approximation of MET in this case is not justified.
Though in the case of a weak gravitational fields (far from sources)
it is absolutely correct. The detailed comparison of MET and GR will
be continued in the next paper.

\section{Conclusion}

The proposed approach of MET gives us natural generalization of GR
in the case of weak gravity. The multidimensional elasticity theory
of strong deformation must be developed to describe strong
gravitational fields.

In general case of weak deformation we get different from GR
equations for space-time dynamics. The strong deformation
theory may give natural physical arguments for including
high curvature terms into Lagrangian of gravitational field.

Multidimensional generalization of classical mechanics and
thermodynamics  (mul\-ti\-di\-men\-si\-o\-nal force, pressure, mass,
temperature,
enthropy) would be discussed separately (see for example
\cite{iv}-\cite{gavr}).

\vspace{2cm}
{\large\bf Aknowlegments}

I am most grateful to R.F. Polishchuk and V.A.Korotky for
useful discussions.

\end{document}